# Using transmission Kikuchi diffraction to characterise α variants in an α+β titanium alloy

V. Tong*, S. Joseph, A. K. Ackerman, D. Dye, and T. B. Britton

Department of Materials, Royal School of Mines, Imperial College London, Prince Consort Road, SW7 2AZ

*Corresponding author: v.tong13@imperial.ac.uk, Telephone: 020 7589 5111

## Abstract

Two phase titanium alloys are important for high performance engineering components, such as aeroengine discs. The microstructures of these alloys are tailored during thermomechanical processing to precisely control phase factions, morphology and crystallographic orientations. In bimodal two phase (α + β) Ti-6Al-2Sn-4Zr-2Mo (Ti-6242) alloys there are often three microstructural lengthscales to consider: large (~10 μm) equiaxed primary α; >200 nm thick plate α with a basketweave morphology; and very fine scaled (>50 nm plate thickness) secondary α that grows between the larger α plates surrounded by retained β. In this work, we utilise high spatial resolution transmission Kikuchi diffraction (TKD, also known as transmission based electron backscatter diffraction, t-EBSD) and scanning electron microscopy (SEM) based forward scattering electron imaging to resolve the structures and orientations of basketweave and secondary α in Ti-6242. We analyse the α variants formed within one prior β grain, and test whether existing theories of habit planes of the phase transformation are upheld. Our analysis is important in understanding both the thermomechanical processing strategy of new bimodal two-phase titanium alloys, as well as the ultimate performance of these alloys in complex loading regimes such as dwell fatigue. Our paper champions the significant increase in spatial resolution afforded using transmission techniques, combined with the ease of SEM based analysis using conventional electron backscatter diffraction (EBSD) systems and forescatter detector (FSD) imaging, to study the nanostructure of real-world engineering alloys.

## 1   Introduction

Titanium alloys are used in mission critical applications, such as within the fan and compressor of an aeroengine. Their high strength to weight ratio and fatigue resistance are exploited to maximum effect when their microstructures are sculpted for the requirements of each application. In disc alloys, such as Ti-6Al-2Sn-4Zr-2Mo, the microstructure is processed to generate a bi-modal microstructure. The two phases at room temperature in these alloys are: (1) a Ti rich α phase, which is made of a hexagonal close packed (HCP) structure; and (2) a Mo rich β phase, which is made of a body centred cubic (BCC) structure. In these two phase bi-modal alloys, thermomechanical processing is used to control the relative size, shape, distribution of the two phases through exploitation of the solid-state α+β (low temperature) to β (high temperature) solid state phase transformation.

Thermomechanical processing is used to control the microstructure. Typically disc alloys contain three major morphologies of α grains, the nomenclature for which is taken from Lütjering & Williams (2007).

- Equiaxed primary α which arise from recrystallisation of deformed α plates;
- Larger α plates with either basketweave or Widmanstätten morphology (>200 nm plate thickness), also termed primary α;
- Fine secondary α plates (<50 nm plate thickness) growing between the basketweave α plates, intermixed with a significant (relative) volume fraction of retained β phase.

The material used in this study is Ti-6Al-2Sn-4Zr-2Mo (Ti-6242) with a bimodal basketweave microstructure. It was processed by rolling in both the β and α+β domains, recrystallised in the α+β domain at 950°C for 5 hours and air-cooled. The alloy was then aged at 593°C for 8 hours to promote secondary α precipitation and then air cooled.

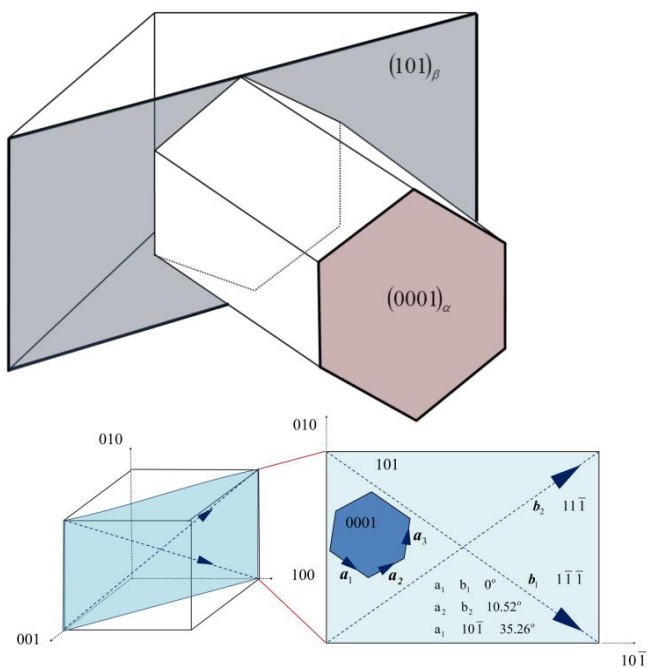

Figure 1: The Burgers orientation relationship between the α (HCP) and β (BCC) phases in a titanium alloy, showing $(0001)_\alpha$ // $\{101\}_\beta$ , $\langle 11\bar{2}0 \rangle_\alpha$ // $\langle 1\bar{1}\bar{1} \rangle_\beta$ and the relationship between HCP <a> type directions and the equivalent BCC directions (adapted from Britton et al. (2015) under an open access CC-BY license).

There exists an orientation relationship between the high temperature β phase and the lower temperature α phase, called the Burgers orientation relationship (BOR), shown in Figure 1. Typically, at high temperatures the β grains grow very large. and α nucleates at β grain boundaries and other defects upon cooling. The orientation of the nucleated α phase is related to the high temperature β phase, such that one $(0001)_\alpha$ is parallel with one $\{101\}_\beta$ and in those planes one <a> type, $\langle 11\bar{2}0 \rangle_\alpha$, direction is parallel with one $\langle 1\bar{1}\bar{1} \rangle_\beta$ direction. In the HCP phase, the <a> direction is a Burgers' vector of the <a> basal and <a> prism slip systems, and in the BCC phase, the Burgers' vector of the typically active slip systems is $\langle 111 \rangle_\beta$ and therefore alignment of these slip systems in the microstructure is important, as it can control the effective slip length.

For each variant, the perfectly aligned <a> direction is called the <$a_1$> direction. In the $(0001)_\alpha$ // $\{101\}_\beta$ plane, there exists a second <a> direction which is reasonably well aligned with a $\langle 11\bar{1} \rangle$ direction (~10.52°, depending on the relative lattice parameters of the α and β phases), and this is called the <$a_2$> direction. The third <a> direction in the $(0001)_\alpha$ is misaligned with respect to Burgers' vectors in the BCC phase, and this is called the <$a_3$> direction.

There are six $\{110\}_\beta$ planes which give rise to six possible $(0001)_\alpha$ orientations. Each $\{110\}_\beta$ plane contains two $\langle 111 \rangle_\beta$ directions, either of which can be aligned to the <$a_1$> direction. These give the 12 possible α variants which can nucleate from a single prior β grain. There have been systematic studies of the preference for different orientations to nucleate, a field of so-called "variant selection". The underlying paradigm of variant selection is that the presence of different daughter variants may be preferable in controlling the ultimate performance of an alloy, in modifying (for example) the effective slip length (Rugg et al. (2007)), propensity of continued basal cleavage during facet fatigue (Sinha et al. (2006)), and the heterogeneity of elastic and thermal expansion of α sub-units (Li et al. (2016)). If few α variants are nucleated on cooling, macrozones (also known as microtextured regions) containing large regions of similar orientation can form in the processed material (Gey et al. (2012)). These act as effective structural units (Rugg et al. (2007)) which facilitate slip through larger volumes of material and can act as crack nucleation sites in dwell fatigue (Uta et al. (2009); Gey et al. (2012)).

The morphology of large basketweave α plates and fine secondary α plates is thought to be controlled by the nucleation and growth of the α phase within the original β grain, though there is some disagreement in the literature as to what the habit plane and growth directions are (Lütjering & Williams (2007); Furuhara et al. (1991)). The broad face habit plane normal and ledged nature of the α and β plates has been studied by Furuhara et al. (1991) using transmission electron microscopy (TEM) and the angular relationships between the α and β plates are reported by Bhattacharyya et al. (2003). The ledge terrace faces are $\{10\bar{1}0\}_\alpha$ // $\{112\}_\beta$ and the ledge step planes are $\{11\bar{2}0\}_\alpha$ // $\{111\}_\beta$. The transverse direction of the ledge is the $[0001]_\alpha$//$\langle 110 \rangle_\beta$ and the broad face habit plane is $\{111\}_\beta$. The broad face habit plane normal in the α crystal frame is not reported in the literature but has been calculated for the present work. In contrast, Lütjering & Williams (2007) report a different broad face habit plane of $\{10\bar{1}0\}_\alpha$ // $\{112\}_\beta$ for the case of Widmanstätten α colonies.

The variants formed in the secondary α may be different to the plate α; having more variants or a different subset of variants in the secondary α may impede fatigue facet nucleation as it would decrease the effective slip length. However, the fine scaled nature of the secondary α make systematic analysis of large numbers of variants, even within one β grain, complex. In a TEM, automated mapping is relatively difficult to perform, excluding the use of newly developed precession electron diffraction techniques (Midgley & Eggeman (2015)), as the majority of analyses of orientation relationships have used well aligned samples pointing along the individual zone axes of the shared $(0001)_\alpha // \{101\}_\beta$ planes (e.g. Savage et al. (2004)).

We have exploited the recent development of the transmission Kikuchi diffraction (TKD) technique to enable this crystallographic analysis to be rapidly performed within a scanning electron microscope (SEM). TKD (also known as transmission electron backscatter diffraction, or t-EBSD) was initially developed by Keller & Geiss (2012) to improve the spatial resolution of orientation measurement in a SEM. In the TKD geometry, the primary electron beam is transmitted through a thin foil specimen such that the majority of electrons pass through the thickness of the foil and some are diffracted to form Kikuchi bands on the phosphor screen. Spatial resolutions of between 8 and 16 nm can be achieved in specimens with thicknesses between 100 and 400 nm (Keller & Geiss (2012); Suzuki (2013)), compared to a few tens of nanometres in conventional electron backscatter diffraction (EBSD) of bulk specimens (Zaefferer (2007); Tong et al. (2015); Chen et al. (2011); Humphreys et al. (1999)). As the majority of the diffracted signal is from the bottom face of the sample closest to the phosphor screen, the presence of grains overlapping through the sample thickness are not generally problematic (Suzuki (2013)). This is in contrast to precession electron diffraction in the TEM, where the signal is generated from the sum of the scattering events throughout the sample thickness during hollow-cone rocking, as indicated in Midgley & Eggeman (2015). A recent detailed review of TKD can be found in Sneddon et al. (2016).

The TKD technique has not been used extensively in titanium, but there are two excellent studies (Sun, Trimby, Yan, et al. (2013); Sun, Trimby, Si, et al. (2013)) to measure relative contributions of twinning and slip to deformation in nanocrystalline commercially pure titanium. The strengthening effect of α precipitates in a near-β titanium alloy has also been studied by Li et al. (2014) and TKD was used to screen for the presence of alpha precipitates before more detailed characterisation using TEM and atom probe tomography.

This paper aims to provide a clear methodology for using orientation and morphology data to distinguish between α variants and the habit planes of α laths. First, forescatter imaging and TKD orientation data are presented. A method for differentiating between α variants is described, and the variant analysis is applied to the TKD map. A method for verifying the habit plane of an α variant is described, and two reported habit planes in the literature are checked for all twelve α variants and the β grain.

## 2 Data acquisition and postprocessing

TKD data was collected on an Auriga FEG-SEM (Zeiss, run at 30kV and a probe current of 10.5 nA measured with a Faraday cup, using a Bruker eFlashHR detector and Esprit 2.1 software [https://www.bruker.com/products/x-ray-diffraction-and-elemental-analysis/eds-wds-ebsd-sem-micro-xrf-and-sem-micro-ct/quantax-ebsd/overview.html]). The stage was tilted to 30° so that the sample plane normal in the TKD holder was parallel to the primary beam. The sample was imaged at a working distance of 1.6 mm and magnification of "100,000×", leading to an image pixel size of 6 nm and TKD step size of 12 nm over the 3 μm × 2.3 μm field of view. The TKD scan took around 40 minutes, during which the sample drifted relative to the beam by around 600nm south and 100nm east.

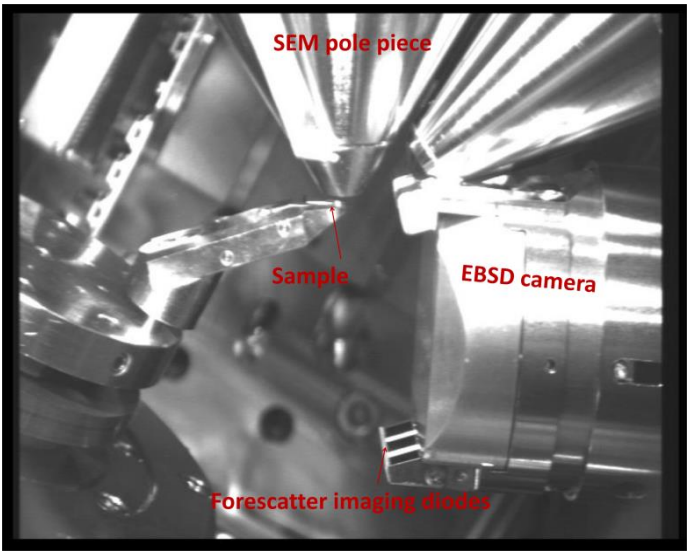

Figure 2: Microscope set up for TKD data acquisition and FSD imaging.

In the eFlashHR detector, three ARGUS forescatter detector (FSD) imaging diodes are mounted at the bottom of the screen (Figure 2) and collect scattered electrons. Electrons hitting each diode independently make up the 'red, green and blue' channels used to create the false colour FSD images.

TKD diffraction patterns were captured with a pattern resolution of 400 × 300 pixels, a pattern centre of $[PC_x, PC_y, PC_z] = [0.45, -0.16, 0.58]$ and the camera was tilted to 4.6° from the horizontal. The pattern centre coordinates are defined according to the Bruker crystal and sample orientation conventions described in Britton et al. (2016).

The EBSD detector was positioned with respect to the sample so that the pattern centre was located slightly off the top of the detector screen. Positioning the detector in this geometry maximises diffraction signal for an untilted sample, but introduces some gnomonic distortions, which are more pronounced near the bottom of the screen. The detector was inserted close to the sample to ensure a wide capture angle; in this experiment the total horizontal capture angle was approximately ~120°. Example diffraction patterns for this configuration and sample are shown in Figure 3.

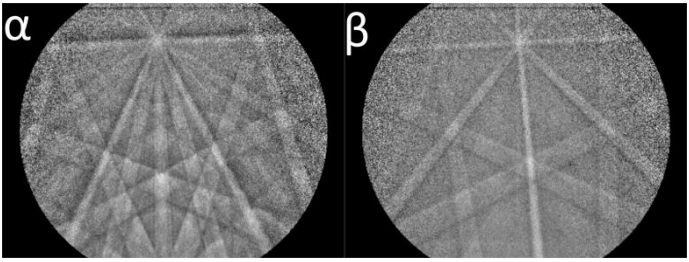

Data was postprocessed within the Bruker Esprit 2.1 software and in MTEX [http://mtex-toolbox.github.io/]. The indexing success rate was >95 %. Wild spike orientations were smoothed out and isolated unindexed points were filled in within the Esprit software, and then exported as a text file. Grain boundaries were identified and pole figures were plotted in MTEX. Pole figures plotted in MTEX were checked against equivalent pole figures plotted in the Esprit software to verify that the same reference frames and orientation descriptions were being used to produce identical pole figures, as care is needed when exporting orientation data into third party software which may use different reference frames and orientation descriptions (Britton et al. (2016)).

## 3   Results

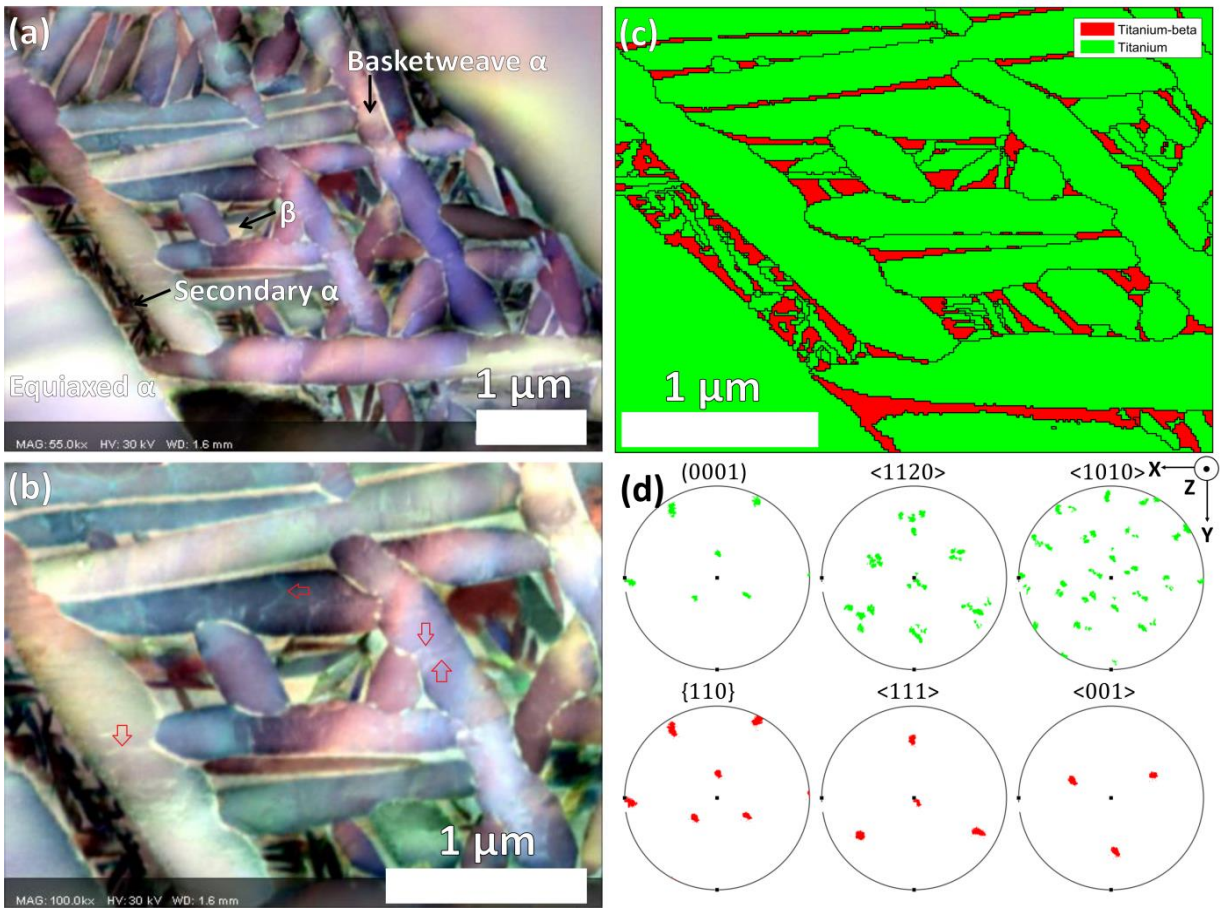

Figure 4 (a) and (b) show colour forescatter ARGUS micrographs of basketweave α plates which have grown into the β matrix between two grain boundary α grains, at lower and higher magnifications respectively. The β phase appears as bright white regions separating α plates, and this contrast is likely dominated by higher Z number elements (such as Mo) segregating to this phase and increasing the amount of scattering (see Supplementary Data A for SEM-based energy-dispersive X-ray spectroscopy data for the two phases).

Contrast within α grains is obtained by collecting signal from electrons hitting different FSD diodes once they have traversed through the foil. The colour contrast has been automatically optimised by the software to reveal different colours within the basketweave plate α and the fine scaled secondary α. The 'rainbow' effect seen within the basketweave α is equivalent to the presence of bend contours in TEM images, and it is shown as a colour variation as the variation of electrons channelling in and channelling out. These are the two dominant orientation imaging contributions (Winkelmann et al. (2007)), causing a different number of electrons to hit the three different diodes used for the red, green and blue false colour channels.

Within these basketweave α plates there are white lines present, typically extending from where one plate intersects a neighbour. These are likely to be dislocations accommodating the misfit strain associated with the growing α plates.

The TKD derived phase map in Figure 4(c) correlates with the higher magnification FSD image shown in Figure 4 (b). This confirms that the brighter phase separating the α plates is the β phase. Note that there has been image drift of between the start of the TKD map and the end of the TKD map, approximately 600nm downwards and 100nm to the right, as evident by the foreshortening of the TKD map image. There is minimal drift in the FSD image as this was captured in less than one minute, and so this represents a more 'true' description of the morphology of the sample. Image drift is a common problem in SEM based orientation mapping, as the TKD map took 41 minutes to capture.

The microstructure of the studied area includes phase data obtained from TKD with grain boundaries overlaid, shown in Figure 4(c). The β (red) phase is confirmed to be the matrix between α (green) laths. The morphology and the relative crystallographic orientations of these two phases are controlled principally by the thermomechanical processing step, and if the strain at room temperature is relatively small (as is the case for this sample), the orientation relationships, morphology and habit planes of this region can be probed using a combination of orientation analysis with TKD and FSD based imaging.

## 3.1  Burgers Orientation Relation and Variant Analysis

In a dual phase titanium alloy, the relationship between the α and β phase should adhere to the BOR. Retained β reflects the high temperature β phase and the α phase will have orientations related to the variants that are formed.

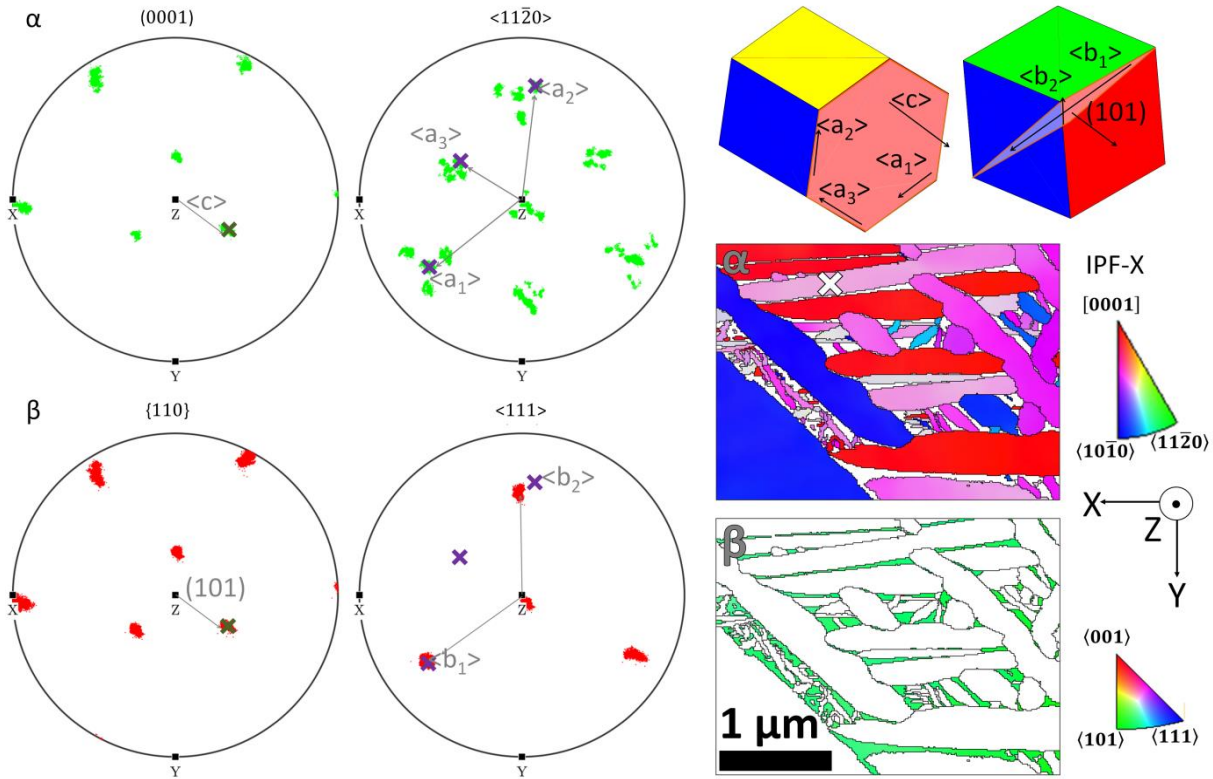

**Figure 5: Method for identifying α variants from pole figures of the α and β grains using stereographic analysis. First the BOR $(0001)_\alpha$ and $(101)_\beta$ are identified as overlapping poles between the α and β phases. Subsequently in the α phase, the <a> directions which are perpendicular to this <c> axis are identified, and in the β phase the <111> directions are identified. Labelling of the <a> direction follows a convention where: <$a_1$> exactly overlaps with a <111>$_\beta$ direction; <$a_2$> is misaligned by ~5° from a <111>$_\beta$ direction; and <$a_3$> does not overlap any <111>$_\beta$ direction (but is closely aligned to <010>$_\beta$).**

Figure 4(d) shows pole figures of the TKD data. For this mapped area, the β phase is a single orientation with the $(111)_\beta$ triad axis pointing out of the page in the z-direction (the unit cell of this β phase is shown in Figure 5).

The α grains grow out of the same parent β matrix forming different daughter variants. All α grains indexed by TKD obey the Burgers orientation relation (BOR), which is described below:

- The six $\{110\}_\beta$ planes in the β phase align with the $\{0001\}_\alpha$ planes of the twelve possible α daughters.
- The <$11\bar{2}0$>$_\alpha$ family contain the <a> type directions in hexagonal close-packed (HCP) titanium. The <$a_1$> and <$a_2$> directions are nearly aligned with the <111>$_\beta$ family, which contain the <$b_1$> and <$b_2$> directions. <$a_1$> is exactly aligned with <$b_1$>, and <$a_2$> is around 10 degrees misoriented from <$b_2$>.
- <$a_3$> is misoriented ~5° from one of three <001>$_\beta$ directions, and is not aligned along any of the <111>$_\beta$ directions.

The final condition is geometrically enforced by the BOR, as <$a_3$> is equiangular from <$a_1$> and <$a_2$> within the basal plane (i.e. they are 120° apart). Similarly, the $[010]_\beta$ direction must lie equiangular from <$b_1$> and <$b_2$> within the $[101]_\beta$ plane, which are both <111>$_\beta$ type directions. It follows that there is a ~5° misorientation between <a$_3$> and <001>$_\beta$ due to the 10° misalignment between <b$_2$> and <a$_2$>.

Figure 5 shows a method of identifying α variants from pole figures.

1. Identify an α grain and its β parent in the TKD map, and identify the points on the pole figure which correspond to points in these grains. The α grain in Figure 5 is marked by a white cross on the IPF-X TKD map.
   a. This information is accessible in the Bruker Esprit 2.1 software in the 'Texture' plotter by hovering over a point on the TKD map. The corresponding pole figure orientations are marked by crosses on the pole figure.
   b. In this case, there is only one prior β grain, so the specific location of the β grain fragment does not need to be identified explicitly as all retained β share the same orientation.
2. Overlay the {0001}$_\alpha$ plane onto the {110}$_\beta$ pole figure. The {110}$_\beta$ plane which matches this orientation is the (101)$_\beta$ plane. These planes are marked out by green crosses on the pole figures in Figure 5.
3. Overlay the <11$\overline{2}$0>$_\alpha$ directions onto the <111>$_\beta$ pole figure. These directions are marked by purple crosses on the pole figures in Figure 5.
   a. There is one <11$\overline{2}$0>$_\alpha$ direction which overlaps with the <111>$_\beta$ direction. These are the <a$_1$> and <b$_1$> directions respectively.
   b. There is one <11$\overline{2}$0>$_\alpha$ direction which is around 10° misoriented from a <111>$_\beta$ direction. These are the <a$_2$> and <b$_2$> directions respectively.
   c. There is a third <11$\overline{2}$0>$_\alpha$ direction which is not oriented close to any <111>$_\beta$ direction. This is the <a$_3$> direction.
4. As with any stereographic projection, the direction of crystallographic vectors (lattice directions and plane normal) are preserved between the stereogram and the two-dimensional surface map, enabling direct projection of these vector directions onto lattice planes and directions in the unit cells, visualised by the hexagon (α) and cube (β) respectively.
5. This process can be repeated for all α grains to create the colour map in Figure 6, where α grains of similar variant are filled in with the same colour.

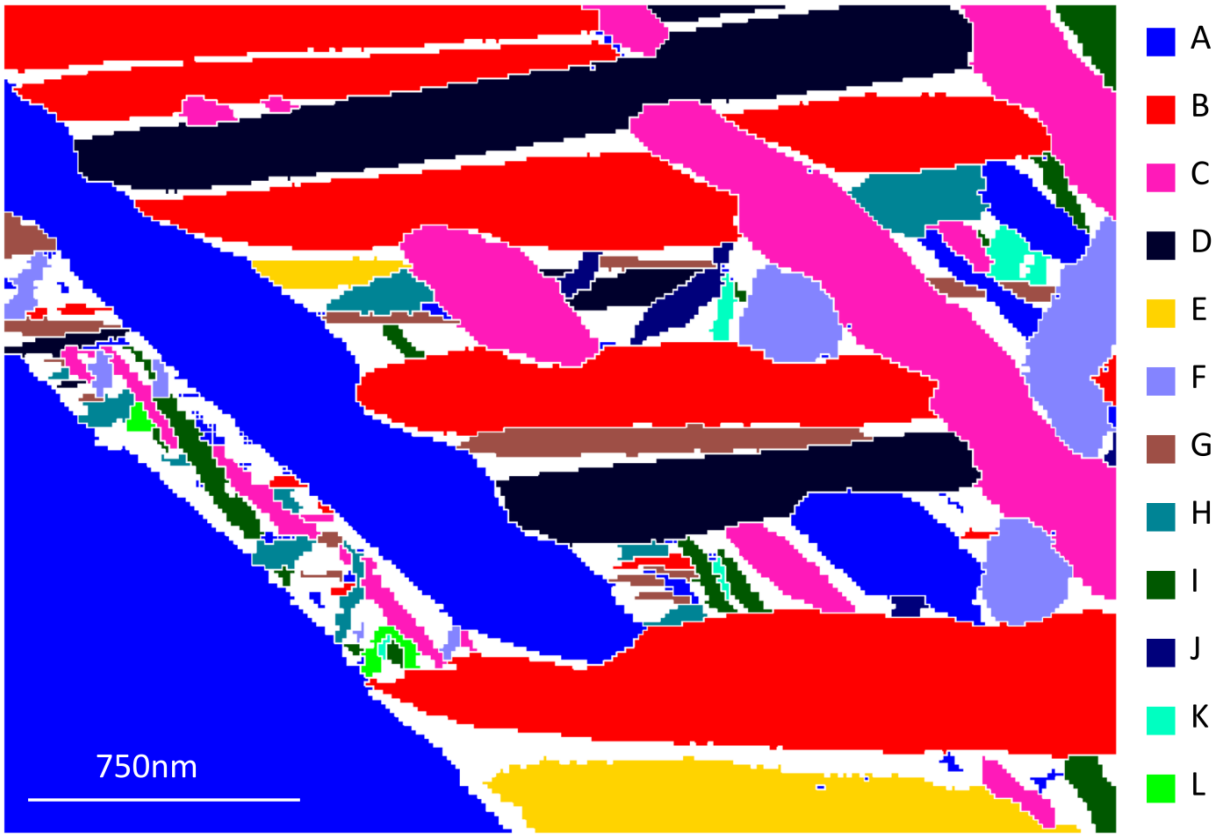

**Figure 6: TKD map of α phase, with different BOR variants filled in with different colours. The variant labelling (A-K) is in order of decreasing area fraction.**

Only six variants (A-F) are present in the larger (>200nm thick) basketweave α plates, whereas all twelve variants are present in the smaller (<100nm thick) secondary α plates. The long axis of the α plates is crystallographic, as all grains of the same variant, both basketweave and secondary α, have their major axis aligned parallel to each other. This is especially obvious for variant C, coloured pink in Figure 6.

## 3.2 Habit planes of α and β plates

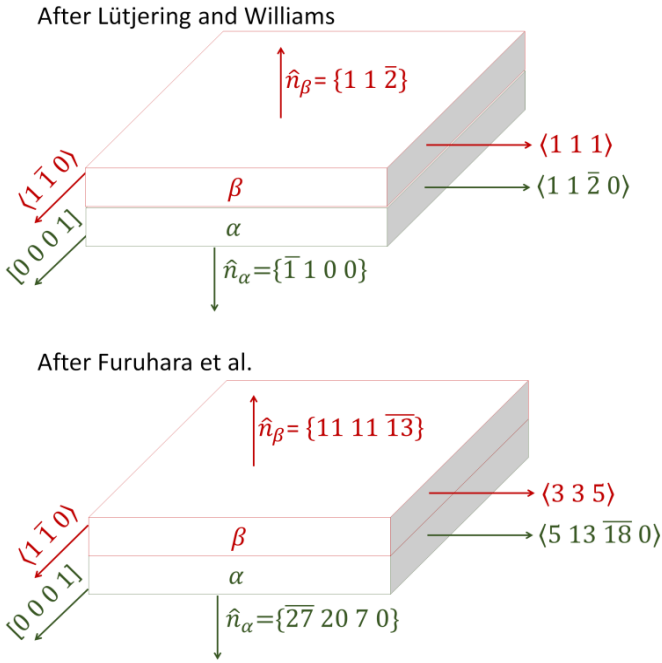

**Figure 7: The growth direction and habit planes of the α and β plates, after Lütjering & Williams (2007) and Furuhara et al. (1991).**

Figure 11: Example pole figure constructions for the habit plane analysis showing variants which follow: the model from Furuhara et al. (1991) (variant A), an ambiguous case (variant B), and the model from Lütjering & Williams (2007) (variant I).Figure 7 shows the habit planes (broad faces) of basketweave α and β plates in Ti-6Al-4V according to Lütjering & Williams (2007) and Furuhara et al. (1991) respectively. Furuhara et al. (1991) provide only the hexagonal habit plane normals with respect to the β plate ($\hat{n}_\beta$ in Figure 7), but since α and β plates are parallel to each other, the habit plane normal for the α plate, $\hat{n}_\alpha$, is unambiguously defined as long as the BOR is obeyed.

Bhattacharyya et al. (2003) describe $\hat{n}_\beta = \{11\ 11\ \overline{13}\}_\beta$ to be misoriented 14.4° from $\{11\overline{2}0\}_\alpha$ about $[0001]_\alpha$. In the Lütjering & Williams (2007) model, the habit plane normal $\hat{n}_\alpha = \{\overline{1}100\}_\alpha$ is misoriented 30° from $\{11\overline{2}0\}_\alpha$ about $[0001]_\alpha$. Therefore, the habit planes in the two models are $30° - 14.4° = 15.6°$ misoriented from each other about $[0001]_\alpha\ //\ \langle1\overline{1}0\rangle_\beta$.

In the present work, $\hat{n}_\alpha$ was calculated by rotating the $(11\overline{2}0)_\alpha$ plane normal by 14.4° about $[0001]_\alpha$ using quaternion rotation of a vector in Cartesian coordinates. The resultant Cartesian vector was transformed back into hexagonal indices and determined to be approximately $\hat{n}_\alpha = \{\overline{27}\ 20\ 7\ 0\}_\alpha$. The cross product between $\hat{n}_\alpha$ and $[0001]_\alpha$ is the second orthogonal in-plane direction of the α plate, $\langle\overline{5}\ \overline{13}\ 18\ 0\rangle_\alpha$. This is also the lattice invariant line direction reported by Furuhara et al. (1991).

In some cases, the boundaries of the α plates are not parallel to each other, and some plate boundaries are not straight but have a zig-zag trace (e.g. the longest plates in variants A and C, shown in Figure 6). This is likely due to the plate morphology being modulated by other microstructural features ahead of the thickening plate. Therefore, in cases where the α plate boundaries are not parallel, the flattest plate face with the least impingement by other grains on its microstructure has been chosen to measure the habit plane normal, as this is likely to be closer to the true habit plane of the plate.

The method of verifying the habit plane normal for each α variant combining spatial mapping of the plate morphology together with the stereographic projection is outlined:

1. Draw a line along the α/β plate boundary on the map, then rotate this line by 90° to construct a vector which represents the projection of the habit plane normal in the sample sectioned plane. There are two edges for each α plate variant – the flatter of the two should be chosen to minimise deviation from the true habit plane via non-uniform α plate growth.

2. Plot the stereographic projection of the candidate habit planes (e.g. $\{\bar{1}12\}_\beta$ // $\{\bar{1}100\}_\alpha$ for Lütjering & Williams (2007) and $\{11\ 11\ \overline{13}\}_\beta$ // $\{\overline{27}\ 20\ 7\ 0\}_\alpha$ for Furuhara et al. (1991)).

3. Take the α or β plate normal projection vector from the forescatter image micrograph and draw this vector onto a stereographic pole figure of each candidate crystallographic habit plane.

4. If the α plate normal lies in the candidate habit plane $\hat{n}_\alpha$, this line should overlap with a plane corresponding to this α plate on the pole figure (i.e. the tested habit plane direction, as represented on the stereogram, is a member of the zone described by the projected trace obtained from Step 1).

5. An equivalent construction can be performed for the β plate normal $\hat{n}_\beta$. As there is only one β grain in this case, all points in the pole figure are used.

6. There can be more than one α or β plane overlapping the habit plane normal construction. This is sometimes true of the β grain with respect to Variant A. As the α and β plates are always parallel to each other, the crystallographic habit planes must also be parallel to each other. Therefore, the correct habit plane must overlap both the α and the β planes in the same orientation position on the pole figure.

7. Note that other overlapping planes may exist which do not lie along the plate normal vectors. Since these planes are not normal to the plate normals, they are irrelevant in the habit plane construction.

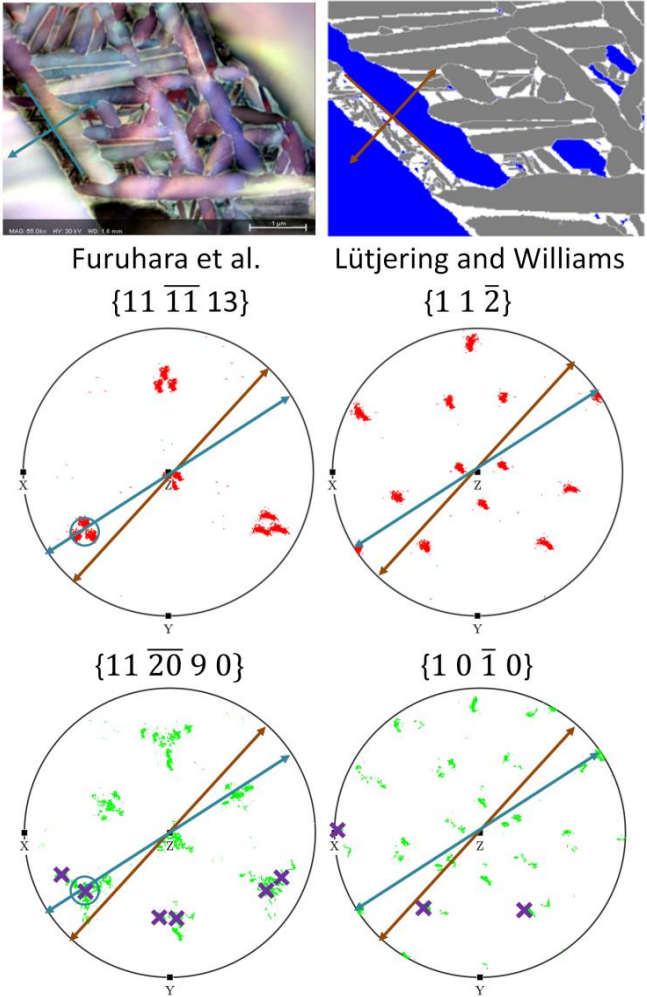

**Figure 8: Habit plane analysis of the secondary α plates. For each variant, traces are extracted from both the FSD image (teal arrows) and orientation map (brown arrows). There is variation due to drift in the TKD mapping step. These are compared against the most likely habit planes, as proposed by Lütjering & Williams (2007) and Furuhara et al. (1991), using a stereographic pole figure based analysis. For the habit plane to be correct, the proposed habit plane must be on the zone described by the projected vector which is perpendicular to the trace of the habit plane extracted from the morphology map. Analysis is performed on potential habit planes for both the α phase (green pole figure) and β phase (red pole figure). As there are multiple α crystal orientations in the pole figure, directions for this variant are highlighted with crosses.**

As there was drift during the TKD scan, the projection vector for the habit normal has been constructed using both the FSD image and the TKD map (i.e. Figure 4(a) and (c) respectively), as there is variation between traces extracted from both. As there is no sample tilt in this TKD configuration, drift present in this scan has only changed the morphology of the features represented in the TKD map (and the orientations remain true). The observation that the orientations are correct and the TKD map is spatially distorted is validated as the habit plane trace analysis is more reasonable when the FSD measured traces are used, shown in Figure 8.

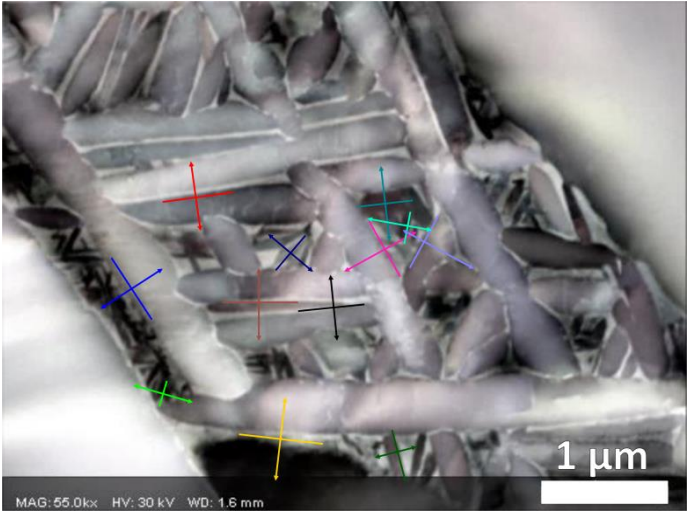

As the FSD based analysis overlaps better with the crystallographic analysis and so the FSD derived habit plane traces have been used to generate Table 1 which compares whether the habit plane belongs to a crystallographic plane set that belongs to one proposed by Furuhara et al. (1991) or Lütjering & Williams (2007). The α plate long axis and plate normals for all twelve variants constructed from the FSD image are shown in Figure 9. From the plate normals, exemplar pole figure constructions outlining the method for habit plane determination are shown in Figure 11, showing a typical construction for each type of result in Table 1. Pole figure constructions for the other nine variants are given in the supplementary data.

| Variant | Furuhara et al. (1991) | | Lütjering & Williams (2007) | |
|---------|------------------------|---|-----------------------------|---|
|         | (α) // {$\overline{27}$ 20 7 0} //(β) | (β) // {$\overline{11}$ 11 13} // (α) | (α) // {1 $\overline{1}$ 0 0} // (β) | (β) // {$\overline{1}$ 1 2} // (α) |
| A | **TRUE** | **TRUE** | FALSE | FALSE |
| B | **TRUE** | **TRUE** | **TRUE** | **TRUE** |
| C | **TRUE** | **TRUE** | **TRUE** | **TRUE** |
| D | **TRUE** | **TRUE** | FALSE | FALSE |
| E | **TRUE** | **TRUE** | FALSE | FALSE |
| F | **TRUE** | **TRUE** | **TRUE** | **TRUE** |
| G | **TRUE** | **TRUE** | FALSE | FALSE |
| H | **TRUE** | **TRUE** | FALSE | FALSE |
| I | FALSE | FALSE | **TRUE** | **TRUE** |
| J | FALSE | FALSE | **TRUE** | **TRUE** |
| K | FALSE | FALSE | **TRUE** | **TRUE** |
| L | FALSE | FALSE | FALSE | FALSE |

This table reveals that (α) // {$\overline{27}$ 20 7 0} and (β) // {11 11 $\overline{13}$} is unambiguously true for variants A-F, corresponding to the variants present in large basketweave α plates. Figure 6 shows that the basketweave and secondary α plates with a variant always lie parallel to each other. Variants C and D (pink and black respectively) show this behaviour particularly clearly.

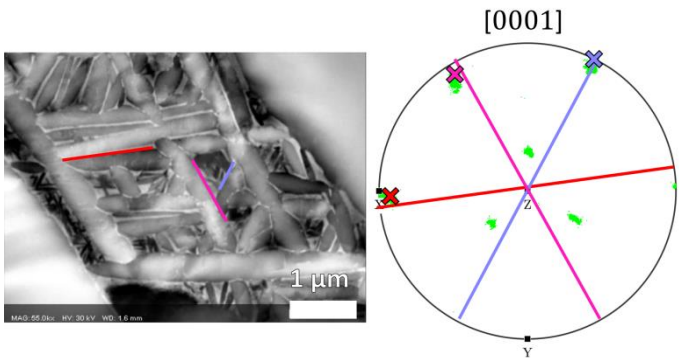

**Figure 10: Variants B (red), C (pink) and F (purple) are sectioned so that the plate long axes are pointing along $[0001]_\alpha$. Therefore, the habit plane for these variants cannot be distinguished between the two models.**

Variants B, C and F also adhere to (α) // $\{1\,\bar{1}\,0\,0\}$ and (β) // $\{\bar{1}\,1\,2\}$ and this is because these variants are sectioned such that the long axis of the sectioned α plate is pointing along $[0001]_\alpha$ (Figure 10). Since $[0001]_\alpha$ is the rotation axis between the two models in Figure 7, the habit plane normal for these variants cannot be distinguished between the two models.

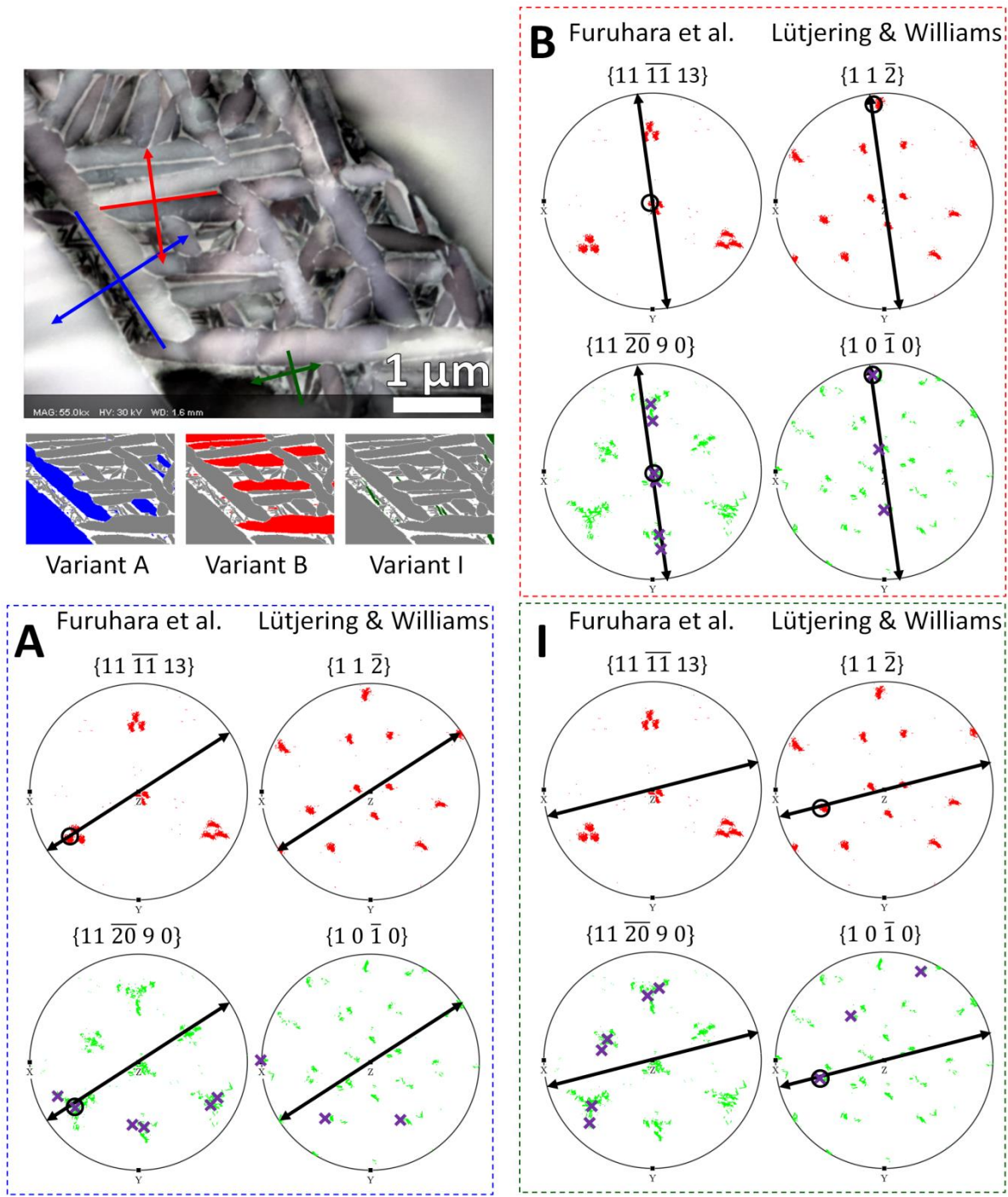

**Figure 11: Example pole figure constructions for the habit plane analysis showing variants which follow: the model from Furuhara et al. (1991) (variant A), an ambiguous case (variant B), and the model from Lütjering & Williams (2007) (variant I).**

The six variants which are present only as secondary α are listed in variants G-L. Two of these variants (G,H) agree with the Furuhara et al. (1991) model whereas the other three variants (I-K) agree with the Lütjering & Williams (2007) model. One variant (L) follows neither model, and has a strange 'U'-shaped morphology in the EBSD map. This seems surprising because in the first six cases (variants A-F) where the larger basketweave α are also present, the secondary α are always parallel to the larger plates. This discrepancy could be due to the increased error in constructing the habit plane normal from the FSD image, as the secondary α plates are very small with respect to the imaging pixel size and necessarily have a lower aspect ratio due to their small size. They are also darker with poor contrast in the FSD image, making plate boundary identification ambiguous. The habit planes in the two models are misoriented only 15.6° from each other, and it is feasible that the measurement uncertainty could be on the same order of magnitude.

## 4   Discussion

Transmission based analysis within the SEM combines ease of use of the SEM, the power of the automated and rapid analysis due to automated diffraction pattern indexing using EBSD, and improved spatial resolution. Using FSD imaging, to capture (in effect) images with high contrast rapidly, reduces the effect of drift on quantitative analysis of morphology in these small regions. The images captured by the FSD images are mostly electron channelling contrast images and contrast seen within the captured images are dominated by changes in the channelling conditions, which are crystal orientation dependent.

However, there is a difference in the interaction volume of the TKD orientation data and the FSD images. Electrons channelling in arise from incoherent scattering events through the sample thickness, and lead to the broad background signal (Winkelmann & Vos (2013)). Electrons can only channel out and form Kikuchi bands if the last scattering event of the electron satisfies the Bragg condition. Therefore, the interaction volume for channelling out is from mostly the bottom surface of the sample. The signal from channelling in is much more intense than the signal from channelling out; this can be seen in raw EBSD patterns which have not been background corrected, where 95% of the total intensity comes from the background.

The FSD diodes collect signal from both channelling in and channelling out of electrons, arising from the entire thickness of the sample, whereas the EBSD Kikuchi band orientation signal comes from channelling out only, and signal is collected only near the bottom surface of the sample. As a result, the spatial resolution is also limited by the presence of multiple features through the sample thickness, and the projection of such features onto the forescatter diodes will affect measurement of the habit plane normal trace.

In this experiment we have experienced a significant challenge in specimen drift. With a 50ms frame capture time for our TKD experiment, the whole map of 256x192 points with a 12nm step size took a total of 41 minutes to capture. The vertical field of view of this map is 2.3 μm, and the drift is estimated to be around 15 nm per minute. Increased use of TKD may drive optimisation of SEMs to reduce sample drift, as manufacturers typically optimise the SEM for short term imaging at high magnification and/or long term stability at a longer length scale (drift is usually about 1 μm per hour).

The TKD map was used to extract orientations of individual variants. These orientations were best combined with the habit planes taken from the FSD imaging, which were subject to less significant drift. Analysis of the habit plates for each of the variants enabled systematic confirmation that the habit plane of the larger basketweave α plates impinging on the β matrix follows the model of Furuhara et al. (1991), where $(\alpha)$ // $\{\overline{27}\ 20\ 7\ 0\}$ and $(\beta)$ // $\{11\ 11\ \overline{13}\}$ and not the model proposed within Lütjering & Williams (2007). Consistent characterisation of the habit planes in this area required us to combine the crystal orientation measurements with the FSD morphology map, thereby reducing the impact of specimen drift when the map is captured using a serial raster. Habit plane analysis of secondary α was less clear due to increased uncertainty in determining the major axis of the α plate and the uncertainty in the habit plane trace where the 'shadow' of an entire three dimensional plate is projected onto a two-dimensional FSD image, as opposed to a 'slice' through the plate as is the case for the larger basketweave laths. This makes it challenging to analyse features that are smaller than the foil thickness with significant confidence, and this has led to uncertainty in the analysis of the smaller alpha laths (G-H).

Table 1 assesses the likelihood of all 12 variants observed within this prior beta grain, and concludes, subject to the assumption that it is the flatter of the two plates which nucleates the alpha plate, that the Furuhara et al. model is true for largest 7 variants (A-G), where the plate faces are less ambiguous due to resolutions issues, whereas the Lütjering and Williams model is potentially only true for 3 cases (and demonstrably false for 4 cases). Of the remaining smaller variants, H-L, the solution is more ambiguous due to the size of the alpha plate and the projection of the plate in this foil. Here we find that either the Furuhara et al. model or the Lütjering and Williams models may be true for different cases.

In this experiment, we have opted to use a satisfactory TKD orientation map to render useful insight into the relative crystallography and variants that are present within our field of view, revealing that the BOR is well adhered to (Figure 5) and that all twelve variants are present (Figure 6). Six variants account for the larger basketweave α plates, and the analysis reveals that these variants are also present within the fine secondary α plates. The presence of as many variants of secondary α as possible, i.e. minimising the extent of variant selection, is important for decreasing the effective slip length.

One of the $\langle 11\overline{2}0 \rangle$ directions in any α variant will be aligned close (<15° misorientation) to a $\langle 11\overline{2}0 \rangle$ direction in eight other variants, due to clustering of orientations from the BOR (Figure 4(d), $\langle 11\overline{2}0 \rangle_{\alpha}$ pole figure). Close alignment of the slip directions makes slip transfer across α plates between any two adjacent variants relatively likely (Guo et al. (2014)). Therefore, a uniform spatial and orientation distribution of all twelve variants is favourable for minimising slip transfer across grain boundaries. Presence of very fine secondary α may also reduce the amplification of stress from a hard-soft grain pair due to the 'rogue grain combination' which is thought to be critical in dwell fatigue (Dunne et al. (2007)).

# 5   Conclusions

A thin foil of dual phase titanium has been studied in the SEM using FSD and TKD to understand the local microstructure and relationships between the two phases. Variant analysis has been performed to identify and distinguish variants in basketweave and fine scale secondary α. The variant map shows that although only six α variants were nucleated in the basketweave plate α, all twelve possible variants were found in the secondary α. The high number of variants nucleated in the secondary α could be favourable for reducing the effective slip length through α and β plates, thereby increasing the strength of the material.

The habit planes for all of the large basketweave plates (variants A-F) are well described by the model reported by Furuhara et al. (1991). The smaller secondary α plates (variants G-L) are in two cases better described by the Furuhara et al. (1991) model (variants G,H), in three other cases the Lütjering & Williams (2007) model (variants I-K), and in one case (variant L) adequately described by neither model. This discrepancy might have arisen from the extremely small size leading to uncertainty in through-thickness projection of the plates in the FSD image, and the relatively lower aspect ratio of secondary α plates, which increases uncertainty when constructing the plane normal.

The habit plane analysis can be ambiguous with some combinations of grain orientation and sectioning plane, present here in variants B, C and F, where due to the sectioning plane used, the measured habit plane seem to fit both models.

We also have demonstrated that a combination of high spatial resolution TKD and FSD based imaging is a valuable approach to focus on combined morphology and orientation relationships for very fine scaled microstructure regions, such as in understanding the habit planes of α plates. The combined approach affords systematic measurement of multiple crystallographic features automatically, using off the shelf hardware and software analysis tools (using TKD) and rapid high contrast imaging (using FSD). This enhances the accuracy of microstructure characterisation promoting a comprehensive understanding of engineering materials.

# 6 Acknowledgments

The authors would like to thank Dr Terry Jun for helpful discussions on variant analysis and providing the supplementary SEM-EDX data, Daniel Goran for advice on using the Bruker Esprit software, and Dr Guang Zeng and Dr Alexander Knowles for help and encouragement with TKD. We would like to thank EPSRC (EP/ K034332/1) for funding on the HexMat Programme Grant (http://www.imperial.ac.uk/hexmat). TBB would like to thank the Royal Academy of Engineering for funding his Research Fellowship.

# 7 Author Contributions

SJ prepared the sample. TBB drove the microscope and supervised the work. VT conducted the analysis and drafted the initial manuscript. DD and AA assisted with understanding of the habit plane analysis. All the authors contributed to reviewing and finalising the manuscript prior to submission.

# 8 Data Statement

The dataset relating to this publication is available at http://doi.org/10.5281/zenodo.439334 as a zipped folder.

The orientation dataset for the TKD map is provided as a text file as-exported from Bruker Esprit 2.1 software.

Compositional data for α and β phases in this alloy are provided in Supplementary Data A.

Pole figure data which is summarised in Table 1 are provided as supplementary figures.